%% file: aaai-submission-latex-2026.tex
\documentclass[letterpaper]{article} 
\usepackage{aaai2026}  
\usepackage{times}  
\usepackage{helvet}  
\usepackage{courier}  
\usepackage[hyphens]{url}  
\usepackage{graphicx} 
\usepackage{natbib}  
\usepackage{caption} 
\usepackage{amssymb}
\usepackage{amsmath}
\usepackage{algorithm}
\usepackage{algorithmic}
\usepackage{multirow}
\usepackage{makecell}
\usepackage{relsize}

\usepackage{graphicx}
\usepackage{subcaption}
\usepackage{array}
\usepackage{tabularx}
\usepackage{diagbox}
\usepackage{graphicx}
\usepackage{pifont}

\usepackage{booktabs} 

\usepackage{times}
\usepackage{helvet}
\usepackage{courier}
\usepackage{xcolor}

\usepackage{newfloat}
\usepackage{listings}
\DeclareCaptionStyle{ruled}{labelfont=normalfont,labelsep=colon,strut=off} 
\floatstyle{ruled}
\newfloat{listing}{tb}{lst}{}
\floatname{listing}{Listing}
\title{Conformal Lesion Segmentation for 3D Medical Images}
\author{
    Binyu Tan\textsuperscript{\rm 1}\equalcontrib, Zhiyuan Wang\textsuperscript{\rm 1}\equalcontrib, Jinhao Duan\textsuperscript{\rm 2}, Kaidi Xu\textsuperscript{\rm 3},\\Heng Tao Shen\textsuperscript{\rm 1,4}, Xiaoshuang Shi\textsuperscript{\rm 1$\dagger$}, Fumin Shen\textsuperscript{\rm 1}\thanks{Corresponding Authors.}
}
\affiliations{
    \textsuperscript{\rm 1}University of Electronic Science and Technology of China\\
    \textsuperscript{\rm 2}University of North Carolina at Chapel Hill\\
    \textsuperscript{\rm 3}City University of Hong Kong\\
    \textsuperscript{\rm 4}Tongji University\\
    \small \texttt{\{binyutan2024, yhzywang, xsshi2013\}@gmail.com} \quad \texttt{jinhao@cs.unc.edu}\\
    \small \texttt{kaidixu@cityu.edu.hk} \quad \texttt{shenhengtao@hotmail.com} \quad \texttt{fshen@uestc.edu.cn}
}

\usepackage{bibentry}

\begin{document}

\maketitle

\begin{abstract}
Medical image segmentation serves as a critical component of precision medicine, enabling accurate localization and delineation of pathological regions, such as lesions. 
However, existing models empirically apply fixed thresholds (e.g., 0.5) to differentiate lesions from the background, offering no statistical guarantees on key metrics such as the false negative rate (FNR). 
This lack of principled risk control undermines their reliable deployment in high-stakes clinical applications, especially in challenging scenarios like 3D lesion segmentation (3D-LS). 
To address this issue, we propose a risk-constrained framework, termed Conformal Lesion Segmentation (CLS), that calibrates data-driven thresholds via conformalization to ensure the test-time FNR remains below a target tolerance $\varepsilon$ under desired risk levels. 
CLS begins by holding out a calibration set to analyze the threshold setting for each sample under the FNR tolerance, drawing on the idea of conformal prediction. 
We define an FNR-specific loss function and identify the critical threshold at which each calibration data point just satisfies the target tolerance. 
Given a user-specified risk level $\alpha$, we then determine the approximate $1-\alpha$ quantile of all the critical thresholds in the calibration set as the test-time confidence threshold.
By conformalizing such critical thresholds, CLS generalizes the statistical regularities observed in the calibration set to new test data, providing rigorous FNR constraint while yielding more precise and reliable segmentations. 
We validate the statistical soundness and predictive performance of CLS on six 3D-LS datasets across five backbone models, and conclude with actionable insights for deploying risk-aware segmentation in clinical practice. 
\end{abstract}


\section{Introduction}
\begin{figure}[!t]
\centering
\includegraphics[width=\linewidth]{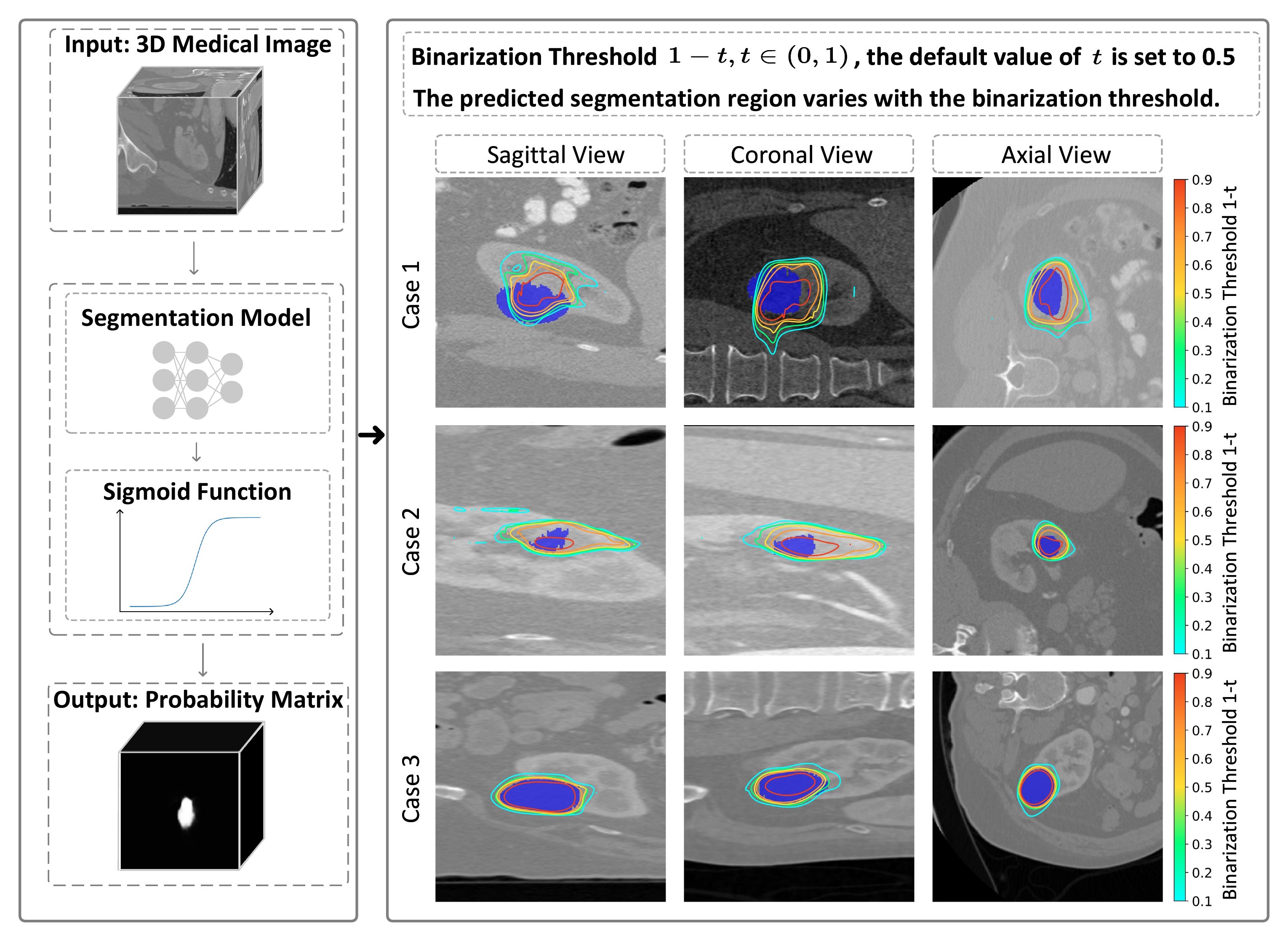}
\caption{Illustration of 3D-LS tasks.}
\label{fig: Illustration}
\end{figure}

Recent progress in deep learning and computer vision~\cite{chen2025survey} has facilitated the development of numerous automated segmentation models for medical imaging modalities, such as computed tomography (CT) and magnetic resonance imaging (MRI)~\cite{moglia2025deep,sun2025foundation}, achieving expert-level performance across various clinical scenarios~\cite{wu2025medical}. 
Despite these gains, current models typically depend on a fixed, heuristic threshold (i.e., 0.5) to delineate target structures from background, lacking statistically valid guarantees for critical safety metrics such as the false negative rate (FNR)~\cite{he2025vista3d}. 
This limitation compromises their trustworthiness in risk-sensitive clinical environments, where robust risk control is essential~\cite{moglia2025deep}. 
In particular, under-segmentation—failing to fully capture lesion boundaries—can result in pathological regions being missed and thus left untreated~\cite{jalalifar2022impact,zhao2025novel}. 
For instance, in early-stage tumor screening, missing lesions smaller than $5$ mm can have serious consequences for patient outcomes, as false negatives directly compromise diagnostic and therapeutic decisions~\cite{korhonen2021breast,luo2023false}. 
These concerns highlight the need for statistically grounded segmentation frameworks, especially in challenging scenarios such as 3D lesion segmentation (3D-LS)~\cite{ni2025advanced,de2025uls23}, 
as illustrated in Figure~\ref{fig: Illustration}, where even small variations in the decision threshold can lead to substantial differences in segmentation outcomes across all three spatial dimensions. 

\begin{figure*}[!t]
\centering
\includegraphics[width=1.0\textwidth]{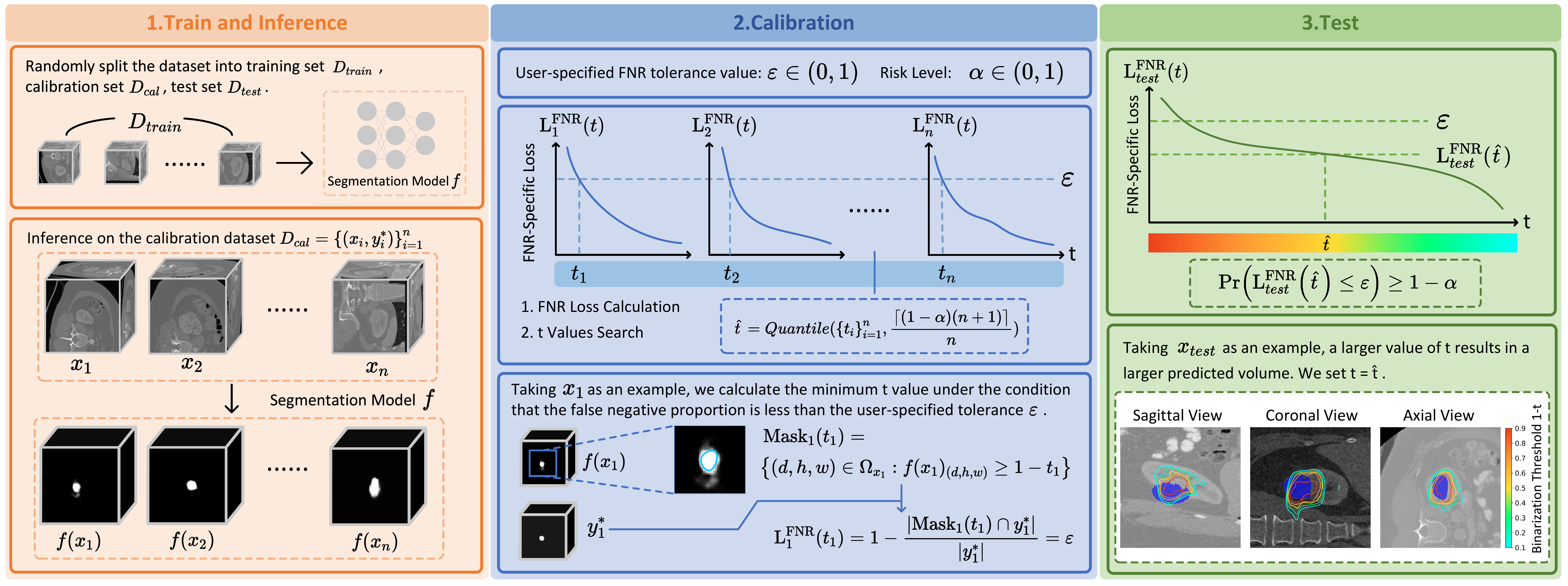} 
\caption{Overview of the CLS framework.}
\label{fig: overview}
\end{figure*}

Split (inductive) conformal prediction (SCP)~\cite{10.1007/3-540-36755-1_29,10.1145/3478535} has recently emerged as a principled solution to address these shortcomings. 
SCP offers distribution-free, model-agnostic guarantees on ground-truth coverage, assuming data exchangeability. 
This framework reserves a held-out calibration set to compute nonconformity scores that quantify the discrepancy between model predictions and ground-truth labels. 
By computing the quantile of these scores at a user-specified risk level $\alpha$ as the test-time selection threshold, SCP then constructs a prediction set for each test data ensuring that the true label is included with probability at least $1 - \alpha$. 
While SCP has demonstrated strong performance in classification settings~\cite{angelopoulos2021uncertainty}, its adaptation to binary image segmentation for FNR control remains a previously under-explored topic.

In this paper, we introduce Conformal Lesion Segmentation (CLS), a novel extension of SCP to image segmentation, which constrains the test-time false negative rate (FNR) to lie below a user-specified tolerance $\varepsilon$, with high probability $1 - \alpha$. 
As shown in Figure~\ref{fig: overview}, CLS builds upon a pretrained segmentation model and focuses on designing a task-specific nonconformity score that reflects lesion-level false negative risk under the given FNR constraint. 
Unlike classification tasks, where the nonconformity score is typically defined as one minus the predicted probability of the true class, segmentation requires careful consideration of how thresholding affects lesion detectability. 
Concretely, for each calibration sample, CLS identifies the maximum decision threshold such that the proportion of false negatives remains below the target tolerance $\varepsilon$. 
This threshold reflects a critical point: increasing it further would begin to exclude true lesion regions, thereby violating the constraint; lowering it might include more lesions but at the cost of increased false positives. 
The collection of these per-sample critical thresholds over the calibration set forms a distribution of nonconformity scores, from which CLS computes the approximate $1 - \alpha$ quantile to determine a statistically calibrated test-time threshold. 
This calibrated threshold ensures, with probability at least $1 - \alpha$, that the FNR on unseen test examples remains within the user-defined tolerance $\varepsilon$.

We evaluate CLS on six 3D-LS benchmarks utilizing five popular 3D medical image segmentation models. 
Empirical results demonstrate that CLS consistently enforces the FNR constraint within the predefined tolerance $\varepsilon$ at the user-specified risk level $\alpha$. 
By rigorously calibrating the test-time threshold, CLS achieves significantly lower FNRs compared to those obtained employing a fixed, heuristic threshold of 0.5 across all datasets. 
Unlike heuristic uncertainty notions, $\alpha$ serves as a statistically rigorous parameter that provides statistically rigorous control over the allowable constraint violation rate. 
Beyond FNR control, we further analyze how different models vary in their predicted region sizes across varying risk levels, offering a practical and interpretable tool for benchmarking uncertainty-aware segmentation models.

Our main contributions are summarized as follows:
\begin{itemize}
    \item We propose Conformal Lesion Segmentation (CLS) that effectively applies SCP to binary segmentation settings. 
    \item We derive novel nonconformity measures from false negative risk-constrained critical thresholds, facilitating statistically rigorous segmentation with FNR control. 
    \item We establish a novel metric for benchmarking model performance specific to uncertainty-aware segmentation. 
\end{itemize}




\section{Related Work}
\noindent \textbf{3D Lesion Segmentation.} 
Recently, specialized 3D segmentation models have been developed to tackle challenges specific to 3D-LS tasks. 
Notable examples include multi-pathway CNNs with CRFs for multiple sclerosis~\cite{saeed2025multi}, two-pathway 3D CNNs that incorporate contextual MRI information for stroke lesions~\cite{bal2024robust}, and lightweight CNN–Transformer hybrids like LW-CTrans for small lesion segmentation~\cite{kuang2025lw}. 
Transformer-based 3D architectures, such as BrainSegFounder~\cite{cox2024brainsegfounder} and MedSAM2~\cite{ma2025medsam2}, integrate multimodal inputs and large-scale pretraining to enhance anatomical representation, while ProLesA-Net~\cite{zaridis2024prolesa} enhances prostate lesion segmentation via multi-channel 3D convolutions. 
Nonetheless, 3D models still inherit the conventional 2D paradigm, applying fixed, heuristic thresholds (e.g., 0.5) to produce binary masks. 
These thresholds are typically uncalibrated and lack statistical guarantees on clinically critical metrics such as the false negative rate (FNR), limiting their reliability in high-stakes medical scenarios.

\begin{figure*}[!t]
    \centering
    \begin{subfigure}{0.35\linewidth}
        \centering
        \includegraphics[width=\linewidth]{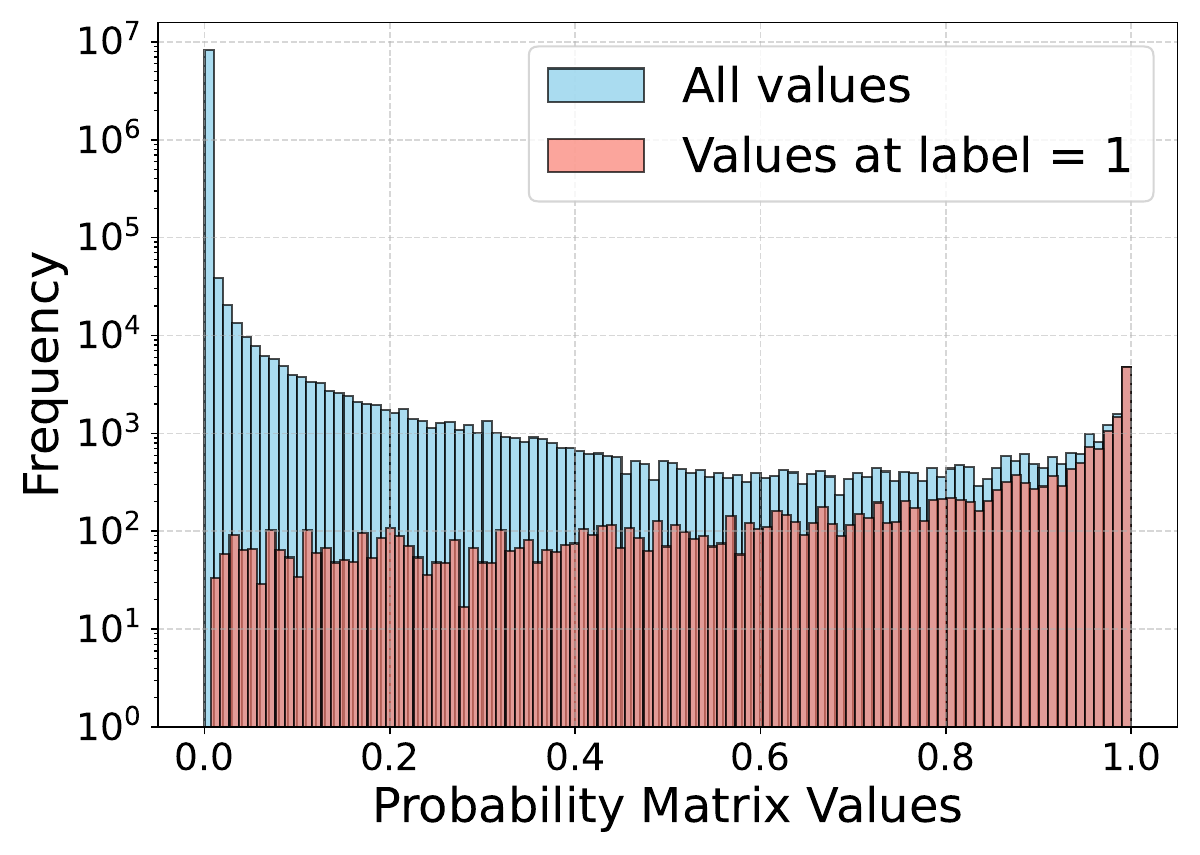}
        \caption{Value Distribution in Probability Matrix.}
        \label{fig: Value Dist. in Prob. Matrix}
    \end{subfigure}
    \hspace{0.1\linewidth}
    \begin{subfigure}{0.35\linewidth}
        \centering
        \includegraphics[width=\linewidth]{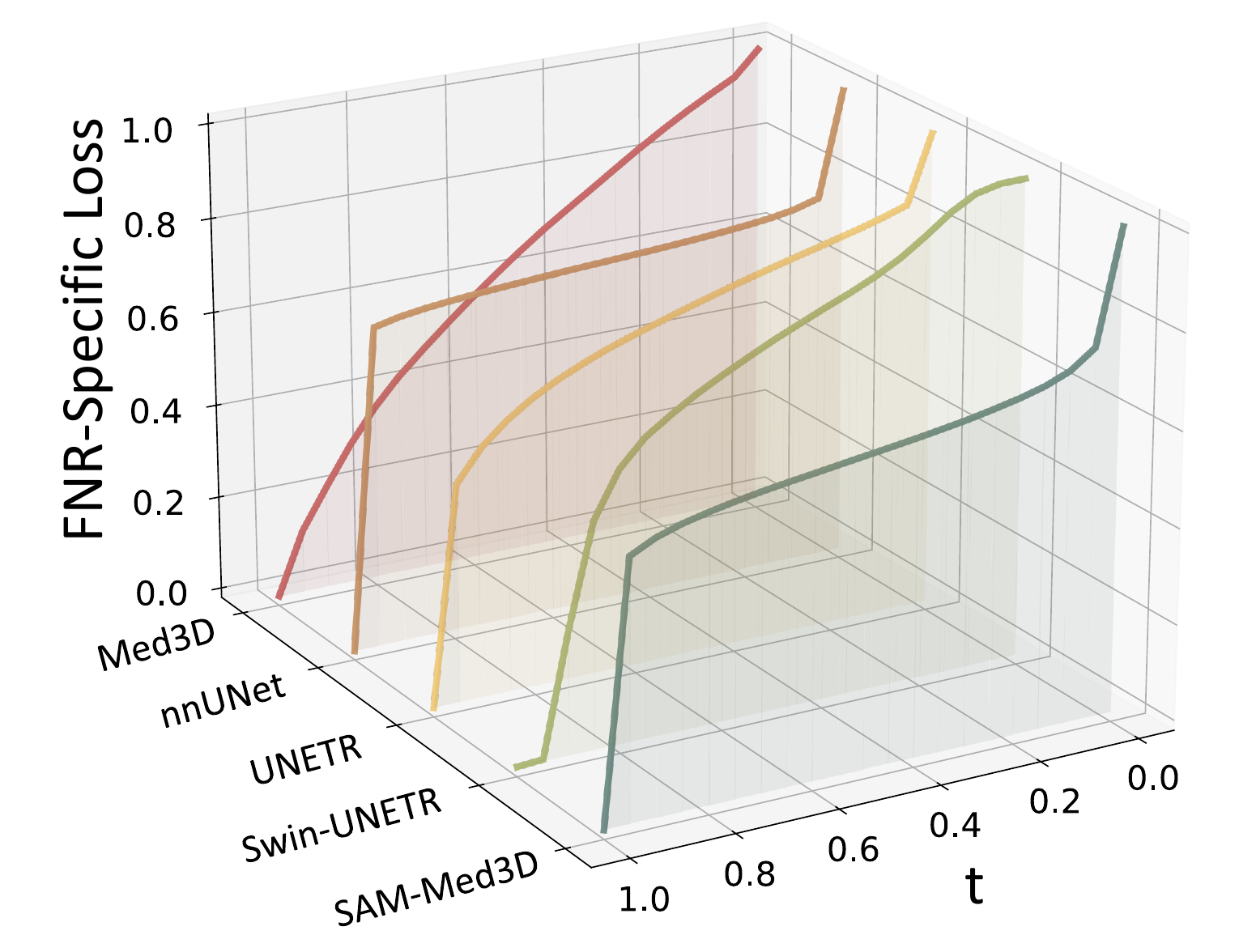}
        \caption{FNR-Specific Loss vs. $t$.}
        \label{fig: FNR-Specific Loss vs. t}
    \end{subfigure}
    \caption{\textbf{(a)} The output probability matrices contain a substantial number of voxels with predicted probabilities below 0.5. Lesion-region probabilities are relatively uniformly distributed, causing nearly half of ground-truth lesion voxels to be missed under a fixed threshold of 0.5. 
    \textbf{(b)} The FNR-specific loss is monotonically non-increasing with respect to $t$.}
    \label{fig: distribution and FNR-Specific Loss}
\end{figure*}

\noindent \textbf{Split Conformal Prediction.} 
SCP is applicable to any pre-trained model to construct sets that are guaranteed to contain the ground truth with a user-specified probability~\cite{angelopoulos2021gentle,wang-etal-2024-conu,wang2025coin}, under the exchangeability condition~\cite{wang2025sconu}.  
Prior studies have effectively applied SCP to image classification scenarios~\cite{liu2025spatial}. 
Under semantic segmentation settings, recent work views segmentation models as a grid of pixel-level classifiers and constructs prediction sets using a pixel-wise SCP approach~\cite{zhi2025seeing}. 
We provide the full conformal procedure under classification settings in the appendix. 
Yet, in binary segmentation tasks like 3D-LS (binary classification), calibrating statistically rigorous prediction sets that validly distinguish foreground lesion regions (class 1) from background (class 0) remains challenging.

\section{Method}
\subsection{Notations and Problem Formulation}
Formally, we begin by partitioning the dataset into three disjoint subsets: a training set $\mathcal{D}_{train}$, a calibration set $\mathcal{D}_{cal}$, and a test set $\mathcal{D}_{test}$. 
Since SCP is compatible with any pretrained model, we first train a segmentation model $f(\cdot)$ using the training set $\mathcal{D}_{train}$. 
Let $x_i\in \mathbb{R}^{D \times H \times W} $ denote a 3D image. 
This model takes $x_i$ as input and produces a confidence map $\hat{y}_i=f(x_{i})\in   \left [ 0,1 \right ]  ^{D \times H \times W}$, where each element represents the predicted probability of the corresponding voxel belonging to a lesion. 
The corresponding ground-truth annotation is given by a binary mask $y_{i}^{*}\in \left \{ 0,1 \right \} ^{D \times H \times W} $, where a value of 1 indicates the presence of pathological tissue at the corresponding location in $x_{i}$. 
Given a decision threshold $t \in [0,1]$, the predicted lesion region is defined as:
\begin{equation}
\mathrm{Mask}_i(t) = \left\{ (d,h,w) \in \Omega_{x_i} : f(x_i)_{(d,h,w)} \ge 1 - t \right\},
\end{equation}
where $\Omega_{x_i} \subset \mathbb{Z}^3$ is the spatial index domain of the 3D image $x_{i}$, $(d,h,w)$ indexes the voxel coordinates, and $f(x_i)_{(d,h,w)}$ is the predicted confidence/probability score. 
Locations with predicted confidence above $1-t$ are classified as foreground (lesion), while the rest are considered background. 

As previously discussed, heuristically setting a fixed decision threshold $1 - t$ (e.g., 0.5) provides no formal guarantee on the false negative risk.
To further illustrate this limitation, we evaluate the pretrained Med3D model~\cite{chen2019med3d} on the KiTS21 dataset~\cite{heller2023kits21}, and present the average value distribution of elements in the output probability matrices in Figure~\ref{fig: Value Dist. in Prob. Matrix}. 
Notably, a substantial portion of predicted probabilities corresponding to ground-truth lesion voxels (i.e., label = 1) fall below 0.5, and these probabilities are distributed relatively uniformly within the lesion regions. 
Such behavior underscores the inadequacy of using a fixed threshold across samples. 
\textbf{Our objective} is to derive a statistically valid threshold $1 - \hat{t}$ on a calibration set $\mathcal{D}_{cal} = \left\{ (x_i, y_i^*) \right\}_{i=1}^n$ such that the test-time false negative rate risk remains below a user-specified tolerance level $\varepsilon$ with high probability, formally formulated as:
\begin{equation}\label{eq: objective}
    \mathrm{Pr}\left( R \left( \hat{t} \right) \leq \varepsilon \right) \geq 1-\alpha,
\end{equation}
where $R \left( \hat{t} \right)$ represents the FNR on fresh test instances (the expectation of the false negative proportion) with the decision threshold of $1-\hat{t}$, and $\alpha$ is a predefined risk level that reflects the maximum acceptable violation/error rate. 

\subsection{Conformal Lesion Segmentation}\label{sec: Conformal Lesion Segmentation}
This section starts with a commonly adopted mild assumption in the SCP framework. 
We then define the FNR-specific loss and introduce a novel nonconformity score tailored for FNR control on the calibration set. 
On this basis, we derive a rigorously calibrated test-time decision threshold and establish its statistical validity. 
Finally, we present the complete workflow of the proposed CLS framework.

\noindent \textbf{\emph{Exchangeable data distribution.}} 
As a foundational yet non-restrictive assumption, we posit that the $n$ calibration samples $\left \{ (x_{i},y_{i}^{*} ) \right \} _{i=1}^{n}$ and each test point $(x_{test}, y_{test}^*)$ in $\mathcal{D}_{test}$ are exchangeable, which underlies the theoretical validity of SCP-based approaches~\cite{angelopoulos2021gentle}. 
Notably, exchangeability is a weaker assumption than independent and identically distributed (i.i.d.) data points. 
We provide a detailed discussion of our assumption in the appendix. 

Given the exchangeability between the given test instance and the calibration data points, the calibration set $\mathcal{D}_{cal}$ can be leveraged as a collection of observed data. 
This enables us to calibrate the segmentation threshold by enforcing an FNR constraint on the calibration set, and to transfer the established statistical guarantees to fresh, unseen data points. 
To align the nonconformity score with the underlying false negative risk, we define a \emph{threshold-dependent FNR-specific loss function} for each calibration data point:
\begin{align}
\mathrm{L}_i^{\text{FNR}}(t) &= 1 - \frac{| \mathrm{Mask}_i(t) \cap y_i^* |}{| y_i^* |}\\&= 1 - 
\frac{\left| \left\{ \substack{(d,h,w) \in \Omega_{x_i} : \\ f(x_i)_{(d,h,w)} \ge 1 - t,\; y^*_{i(d,h,w)} = 1} \right\} \right|}
     {\left| \left\{ (d,h,w) \in \Omega_{x_i} : y^*_{i(d,h,w)} = 1 \right\} \right|} \nonumber,
\end{align}
where $y^*_{i(d,h,w)}$ is the ground-truth label at voxel $(d,h,w)$ and the loss reflects the proportion of false negatives among all ground-truth lesion voxels, evaluated at threshold $1-t$. 
A lower loss indicates that a larger fraction of the ground-truth lesion voxels are successfully identified by the model. 

The monotonicity of the FNR-specific loss (i.e., the false negative proportion of each sample under a given decision threshold $1 - t$) with respect to $t$ is immediate from its definition, as also illustrated in Figure~\ref{fig: FNR-Specific Loss vs. t}. 
Leveraging this property, we then develop the \emph{nonconformity score} as
\begin{equation}
    t_i=\inf\left \{ t:\forall t'\ge t, \mathrm{L}_i^{\text{FNR}}(t')\le \varepsilon  \right \},
\end{equation}
which represents the lowest feasible decision threshold for each calibration sample under which the false negative proportion remains within the specified tolerance $\varepsilon$. 
This critical point $t_i$ minimizes the predicted lesion area $\mathrm{Mask}_i(t_i)$ to enhance precision subject to the risk constraint. 
Subsequently, we sort all nonconformity scores $\{t_i\}_{i=1}^n$ on the calibration set in ascending order such that $t_1\le \cdots  \le t_n\ $, and compute their $\frac{\left \lceil (1-\alpha )(1+n) \right \rceil }{n} $ quantile:
\begin{align}\label{eq: t-hat}
    \hat{t}=\inf\left \{ t:\frac{\left |\{ i:t_i\le t \}\right | }{n} \ge \frac{\lceil (1-\alpha )(1+n)  \rceil}{ n}  \right \}. 
\end{align}

\noindent \textbf{Theorem 1} (Statistically rigorous FNR constraint)\textbf{.} 
\textit{Suppose the given test instance $(x_{test},y_{test}^*)$ and $(x_i,y_i^*)_{i=1,\cdots,n}$ are exchangeable, we employ $\hat{t}$ as the test-time decision threshold and the resulting predicted lesion region is $\mathrm{Mask}_{test}(\hat{t})$. 
Then the false negative proportion $\mathrm{L}^{\text{FNR}}_{test} \left( \hat{t} \right)$ satisfies} 
\begin{equation}
    \mathrm{Pr}\left( \mathrm{L}^{\text{FNR}}_{test} \left( \hat{t} \right) \leq \varepsilon \right) \geq 1-\alpha.
\end{equation}

This is the same property of risk control as Eq.~\eqref{eq: objective}. 
Below, we establish its statistical rigor. 

\noindent \textit{Proof of Theorem 1.} 
Under the condition that $s_{i=1, \cdots, n}$ are in ascending order, $\hat{t}$ can be reformulated as
\begin{equation}
    \hat{t} = t_{n\cdot \frac{\left \lceil (1-\alpha )(1+n) \right \rceil }{n}}=t_{\left \lceil (1-\alpha )(1+n) \right \rceil }. 
\end{equation}
By the definition of $t_i$, if $\mathrm{L}^{\text{FNR}}_{test} \left( \hat{t} \right) \leq \varepsilon$, it can obtain
\begin{equation}
    \hat{t} \geq t_{test}. 
\end{equation}
Since $(x_1, y_1^*), \cdots, (x_n, y_n^*), (x_{test}, y_{test}^*)$ are supposed to be exchangeable, it has
\begin{equation}
     \mathrm{Pr}\left( t_{test} \leq t_i \right) =\frac{i}{n+1}. 
\end{equation}
Finally, it can obtain
\begin{equation}
    \begin{split}
        \mathrm{Pr}\left( \mathrm{L}^{\text{FNR}}_{test} \left( \hat{t} \right) \leq \varepsilon \right) &= \mathrm{Pr}\left( \hat{t} \geq t_{test}\right)\\
    &=\mathrm{Pr}\left( t_{\left \lceil (1-\alpha )(1+n) \right \rceil } \geq t_{test}\right)\\
    &=\frac{\left \lceil (1-\alpha )(1+n) \right \rceil}{n+1}\\
    & \geq 1-\alpha
    \end{split}.
\end{equation}
This completes the proof of Theorem 1 and establishes the statistical validity of the test-time decision threshold. 




\noindent \textbf{Workflow of CLS.} 
Given a target FNR tolerance $\varepsilon$, we compute nonconformity scores $\{t_i\}_{i=1}^{n}$ on the calibration set to determine the threshold settings under the risk constraint. 
We proceed to compute the approximate $1-\alpha$ quantile $\hat{t}$ of these critical scores, corresponding to the user-specified risk level $\alpha$. 
At test time, $\hat{t}$ is employed as the decision threshold for unseen instances. 
With probability at least $1 - \alpha$, the FNR on the test set is guaranteed to remain below $\varepsilon$. 
We provide the corresponding pseudocode in the appendix.


\begin{figure*}[!t]
  \centering

  \begin{subfigure}[b]{0.33\textwidth}
    \includegraphics[width=\textwidth]{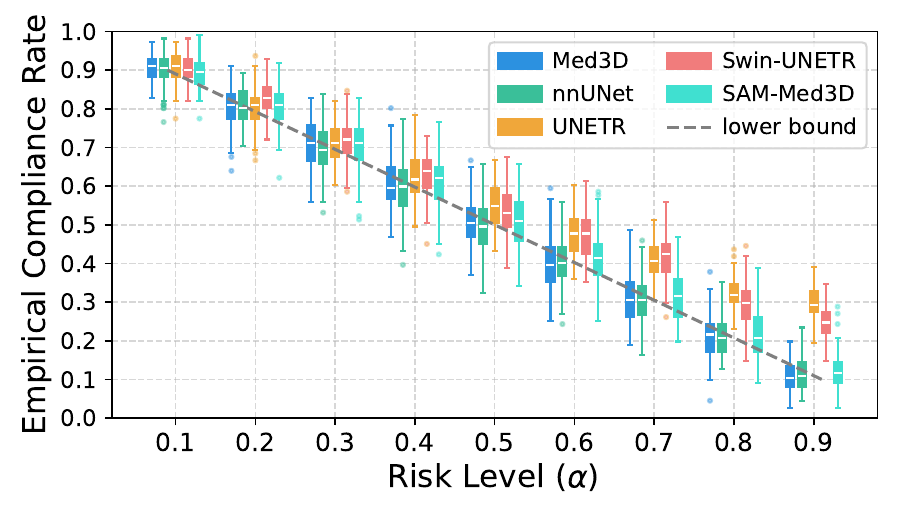}
    \caption{KiTS21.}
    \label{fig:ECR results of KITS21}
  \end{subfigure}
  \hfill
  \begin{subfigure}[b]{0.33\textwidth}
    \includegraphics[width=\textwidth]{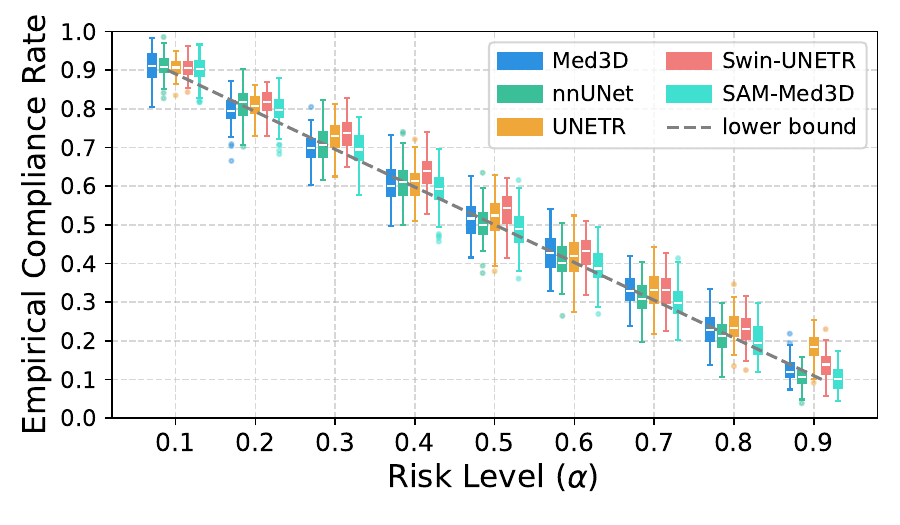}
    \caption{LiTS.}
    \label{fig:ECR results of LITS}
  \end{subfigure}
  \hfill
  \begin{subfigure}[b]{0.33\textwidth}
    \includegraphics[width=\textwidth]{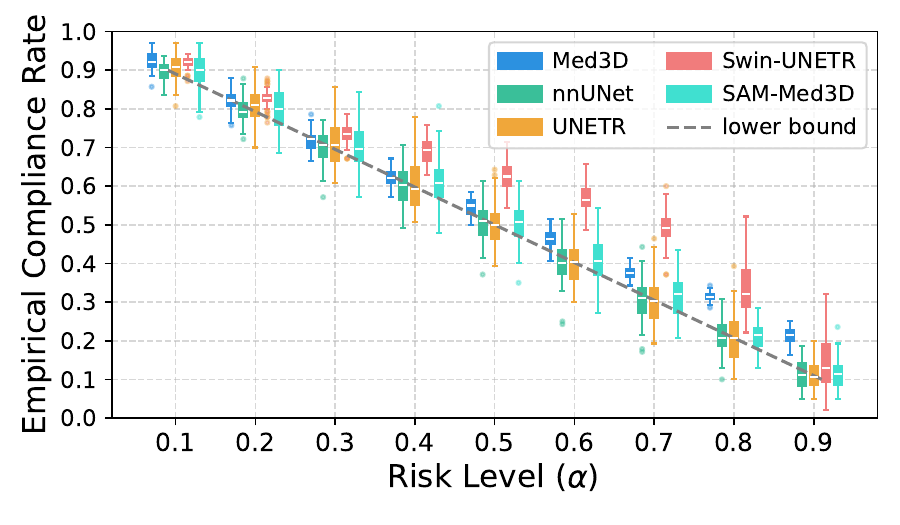}
    \caption{NIH-LN ABD.}
    \label{fig:ECR results of NIH-LN ABD}
  \end{subfigure}

  \vspace{1em}

  \begin{subfigure}[b]{0.33\textwidth}
    \includegraphics[width=\textwidth]{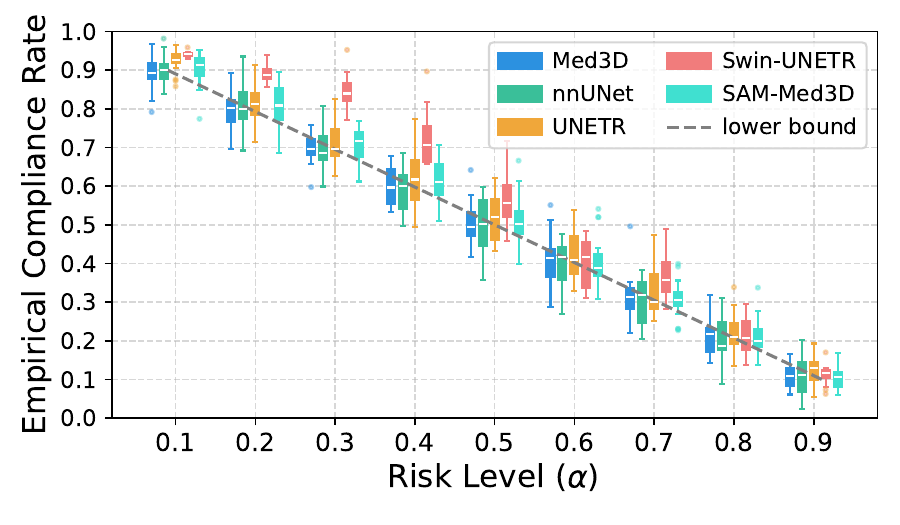}
    \caption{LIDC-IDRI.}
    \label{fig:ECR results of LIDC-IDRI}
  \end{subfigure}
  \hfill
  \begin{subfigure}[b]{0.33\textwidth}
    \includegraphics[width=\textwidth]{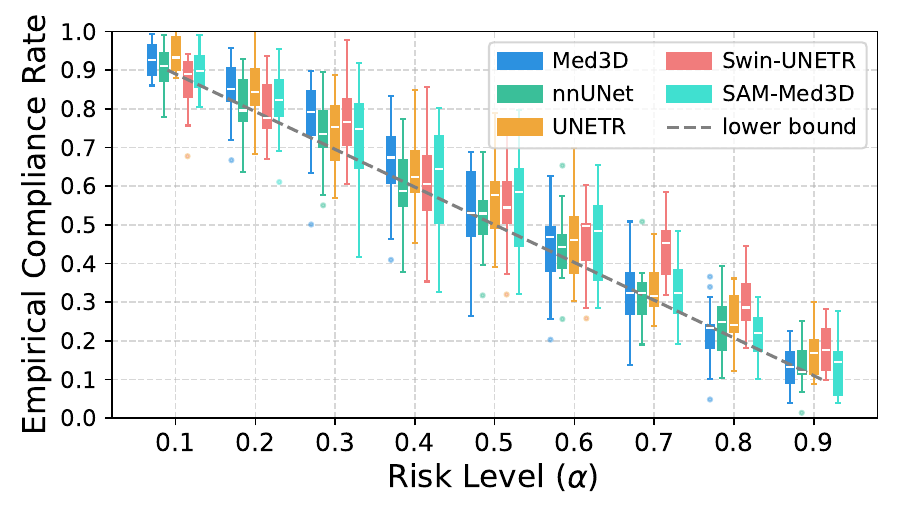}
    \caption{MDSC-Colon.}
    \label{fig:ECR results of MDSC-Colon}
  \end{subfigure}
  \hfill
  \begin{subfigure}[b]{0.33\textwidth}
    \includegraphics[width=\textwidth]{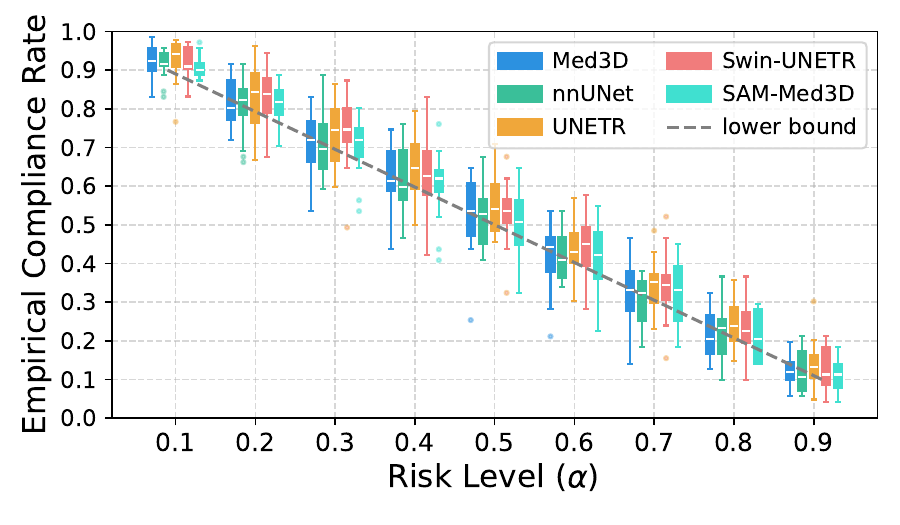}
    \caption{MDSC-Pancreas.}
    \label{fig:ECR results of MDSC-Pancreas}
  \end{subfigure}

  \caption{Test-time ECR results on six datasets utilizing five pretrained segmentation models. }
  \label{fig:ECR}
\end{figure*}

\input{tables/average-FNR-specific-loss}

\section{Experiments}
\subsection{Experimental Settings}
\noindent \textbf{Datasets.} We consider six fully annotated 3D medical segmentation datasets from the ULS23 Segmentation Challenge \cite{de2025uls23}, each covering a different anatomical region or organ system: KiTS21 \cite{heller2023kits21}, LiTS \cite{bilic2023liver}, NIH-LN ABD \cite{roth2014new}, LIDC-IDRI \cite{armato2011lung}, MDSC-Colon \cite{antonelli2022medical}, and MDSC-Pancreas \cite{antonelli2022medical}. 
More details of the utilized datasets are provided in the appendix. 

\input{tables/average-ECR}

\noindent \textbf{Backbone Models.} 
We adopt five popular 3D medical image segmentation models with diverse architectural designs, each fine-tuned to achieve a comparable number of parameters: Med3D\cite{chen2019med3d}, nnUNet\cite{isensee2021nnu}, UNETR\cite{hatamizadeh2022unetr}, Swin-UNETR\cite{hatamizadeh2021swin}, and SAM-Med3D\cite{wang2023sam}. More details are provided in the appendix.


\noindent \textbf{Evaluation Metrics.} 
We check whether the \emph{\textbf{empirical compliance rate}} (ECR)~\cite{angelopoulos2021gentle}, defined as the proportion of test samples whose FNR-specific loss is controlled below $\varepsilon$, exceeds $1-\alpha$. 
Beyond FNR control, we also emphasize spatial precision by encouraging compact lesion predictions, as smaller predicted regions are generally more accurate and clinically preferable.
We introduce \textbf{\emph{prediction compactness}} (PC), defined as the ratio of the number of predicted lesion voxels to the total number of voxels in the input image. 
We evaluate PC under the constraint that the FNR remains within the specified tolerance $\varepsilon$, and examine how PC varies with different risk levels $\alpha$ and across models. 
While the denominator in PC can alternatively be formulated using the ground-truth lesion volume, such a choice only introduces a constant scaling factor per sample and does not affect the relative ranking of models. 
Thus, PC provides a robust and interpretable measure for comparing the spatial efficiency of lesion predictions under risk constraints.

\subsection{Empirical Evaluations}
We set the split ratio between the calibration set and the test set to 0.5 by default. 
Each experimental group is evaluated over 100 random splits of the calibration and test samples. 

\noindent \textbf{Statistical Validity of CLS.} 
We begin by demonstrating the statistical rigor of Theorem 1. 
As illustrated in Figure~\ref{fig:ECR}, utilizing each decision threshold calibrated by Eq.~\eqref{eq: t-hat} effectively constrains the test-time ECR metric under various user-specified risk levels on all six 3D-LS datasets across five pretrained segmentation models. 
Notably, we expect the ECR results on the test set to approach but remain above the $1-\alpha$ lower bound~\cite{angelopoulos2021gentle,wang2025sconu} under the strict assumption of data exchangeability. 
In 100 test runs, we occasionally observe that the ECR falls below the theoretical lower bound of $1 - \alpha$. While the guarantee provided by the SCP framework is statistically rigorous~\cite{ye2024benchmarking,wang2025sample}, minor violations can occur in practice due to finite-sample variability.

\noindent \textbf{Comparison with Heuristic Thresholding.} 
We conduct a comprehensive empirical study to compare the performance of CLS with conventional heuristic thresholding in test-time risk-sensitive segmentation. 
As presented in Table~\ref{tab:ablation thresholding mean}, under a fixed threshold of 0.5, all five pretrained segmentation models exhibit substantially high test-time FNRs on the KiTS21 dataset, ranging from 0.4935 (SAM-Med3D) to 0.5804 (UNETR). 
These results underscore the limitations of heuristic decision thresholds, which fail to adapt to distributional uncertainty and lead to frequent false negatives—an issue particularly critical in clinical settings. 
By contrast, CLS develops nonconformity scores based on a tailored FNR-specific loss, enabling the derivation of statistically rigorous thresholds under a user-specified risk level (e.g., $\alpha = 0.2$). This calibration leads to substantial and consistent FNR reductions across all models and tolerance levels ($\varepsilon$). 
For instance, at a moderate tolerance of $\varepsilon = 0.2$, CLS reduces the FNR of Med3D from 0.5338 to 0.1032, achieving an $80.7\%$ relative reduction. Similarly, significant improvements are observed for nnUNet (from 0.5122 to 0.0933), UNETR (from 0.5804 to 0.1145), Swin-UNETR (from 0.5642 to 0.1070), and SAM-Med3D (from 0.4935 to 0.0956).
Furthermore, CLS consistently satisfies the specified FNR tolerance across a wide range of risk levels ($\varepsilon \in \{0.1, 0.2, 0.3, 0.4, 0.5\}$), demonstrating both statistical rigor and adaptability. Notably, even under the strictest constraint ($\varepsilon = 0.1$), FNRs remain below target across all models (e.g., 0.0466 for Swin-UNETR), while relaxed tolerances are met with less conservative thresholds, offering a flexible trade-off between coverage and safety. 

\begin{figure}[!t]
\centering
\includegraphics[width=\linewidth]{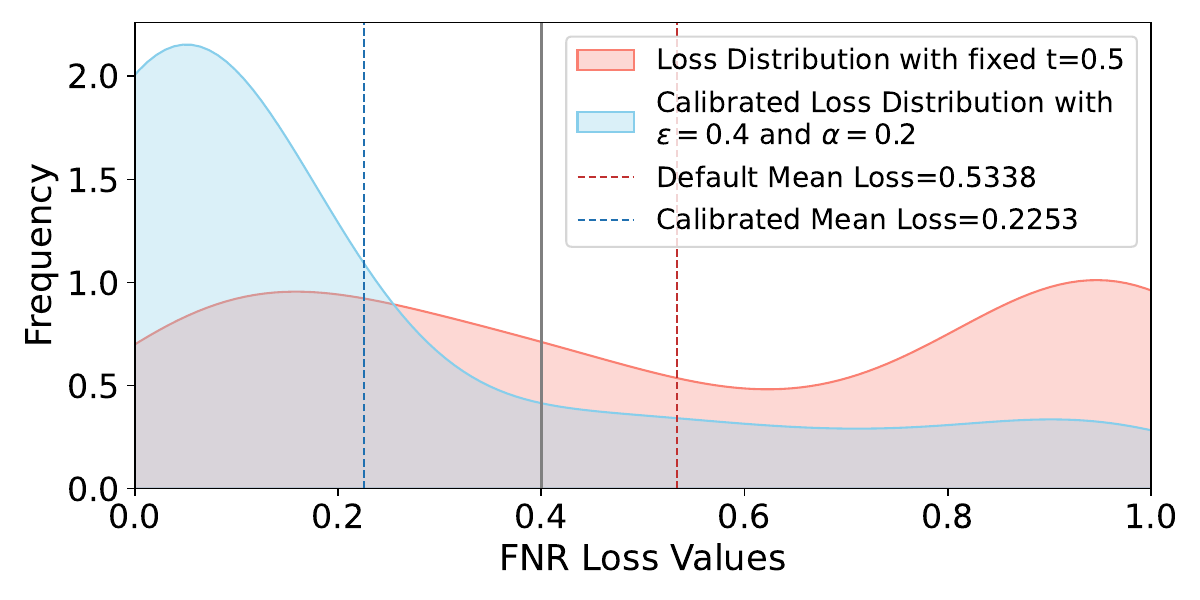}
\caption{Distribution of FNR-specific loss before (fixed threshold at $0.5$) and after CLS calibration ($\varepsilon = 0.4$, $\alpha = 0.2$), illustrating the reduction in average loss and variance.}
\label{fig: FNR Frequency}
\end{figure}

To understand the underlying calibration behavior from a probabilistic lens, we visualize the distribution of FNR-specific loss values before and after applying CLS calibration. 
As illustrated in Figure~\ref{fig: FNR Frequency}, the loss distribution under a fixed threshold (red) is skewed rightward with a high mean of 0.5338, indicating systematic lesion under-segmentation. After CLS calibration with $\varepsilon = 0.4$ and $\alpha = 0.2$ (blue), the distribution shifts sharply leftward, with the mean reduced to 0.2253. This reflects CLS’s ability to tightly control false negatives by adjusting thresholds based on sample-specific risk, resulting in globally improved test-time performance. 

We further examine the generalization capability of CLS across multiple datasets.
As shown in Table~\ref{tab:ablation thresholding p}, when using a fixed threshold, ECR values remain low across six 3D lesion segmentation (3D-LS) datasets, with several models performing below 0.40 (e.g., 0.3029 for Med3D on MDSC-Colon, 0.3134 for Swin-UNETR on LiTS).
After applying CLS calibration at a risk level of $\alpha = 0.2$, ECR improves substantially and consistently, achieving gains of over +0.45 absolute improvement on average. For example, the ECR of Med3D on LIDC-IDRI increases from 0.4783 to 0.8137, while Swin-UNETR on KiTS21 jumps from 0.4081 to 0.8289, indicating a near doubling in lesion coverage. These results confirm that CLS not only enforces failure rate constraints but also enhances practical segmentation quality. 

Importantly, gains are model-agnostic and task-invariant, demonstrating the robustness of CLS across both CNN-based and Transformer-based backbones, as well as across diverse organ and modality domains.
By jointly controlling FNR and maximizing coverage, CLS offers a principled and practically effective framework for deploying segmentation models under clinically meaningful risk constraints.

\begin{figure*}[!t]
  \centering

  \begin{subfigure}[b]{0.32\textwidth}
    \includegraphics[width=\textwidth]{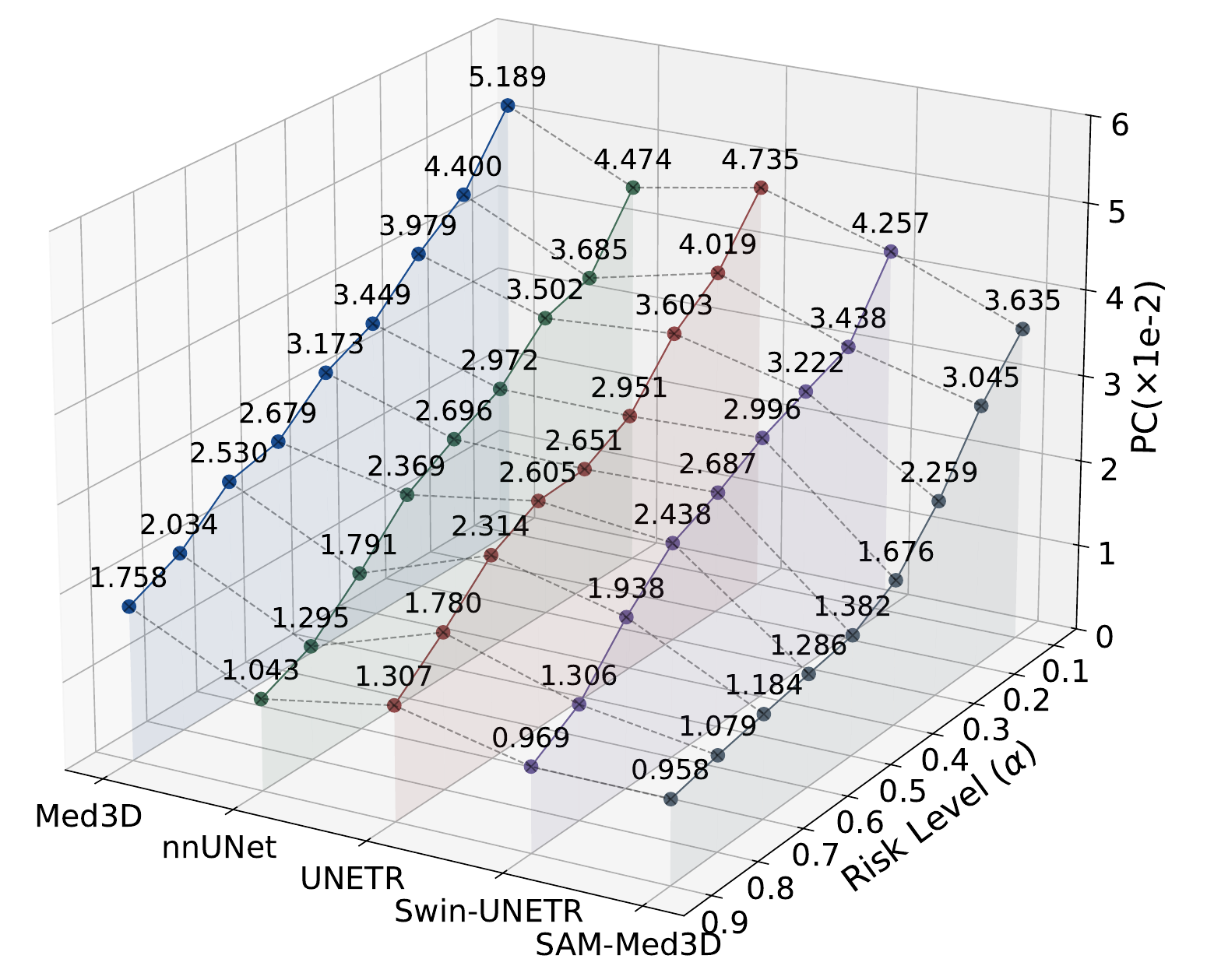}
    \caption{KiTS21.}
    \label{fig:APLVR results of KITS21}
  \end{subfigure}
  \hfill
  \begin{subfigure}[b]{0.32\textwidth}
    \includegraphics[width=\textwidth]{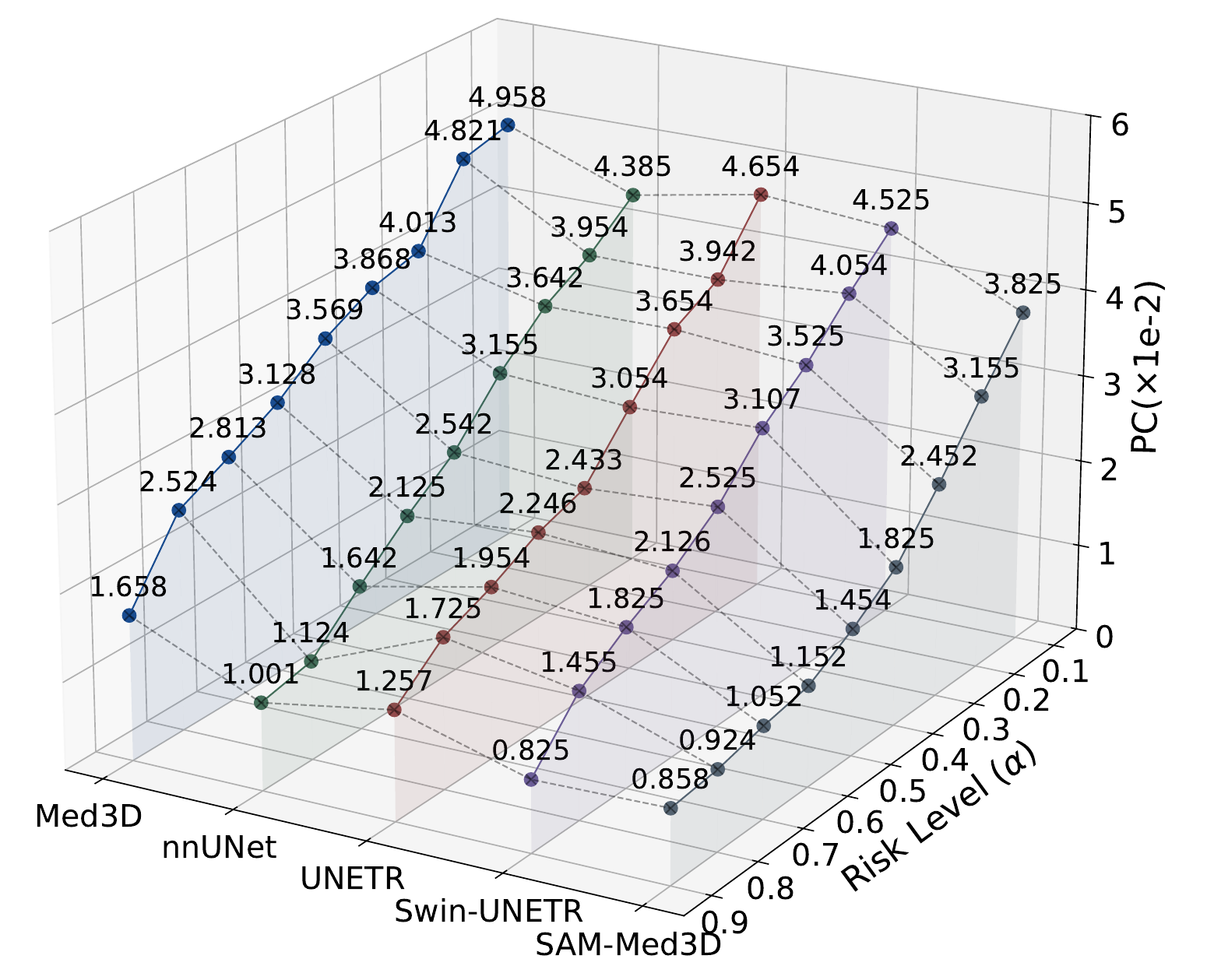}
    \caption{LiTS.}
    \label{fig:APLVR results of LITS}
  \end{subfigure}
  \hfill
  \begin{subfigure}[b]{0.32\textwidth}
    \includegraphics[width=\textwidth]{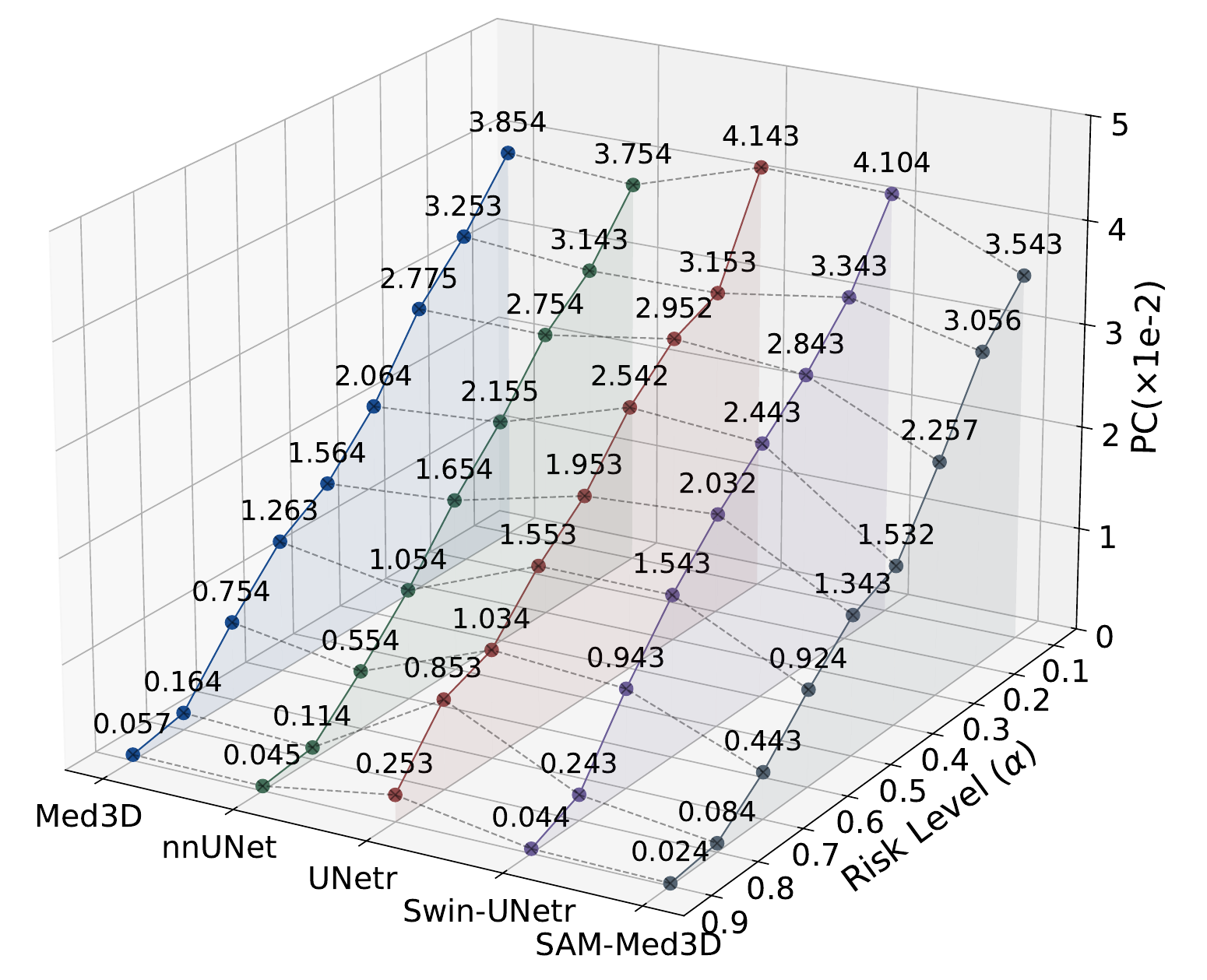}
    \caption{NIH-LN ABD.}
    \label{fig:APLVR results of NIH-LN ABD}
  \end{subfigure}


  \begin{subfigure}[b]{0.32\textwidth}
    \includegraphics[width=\textwidth]{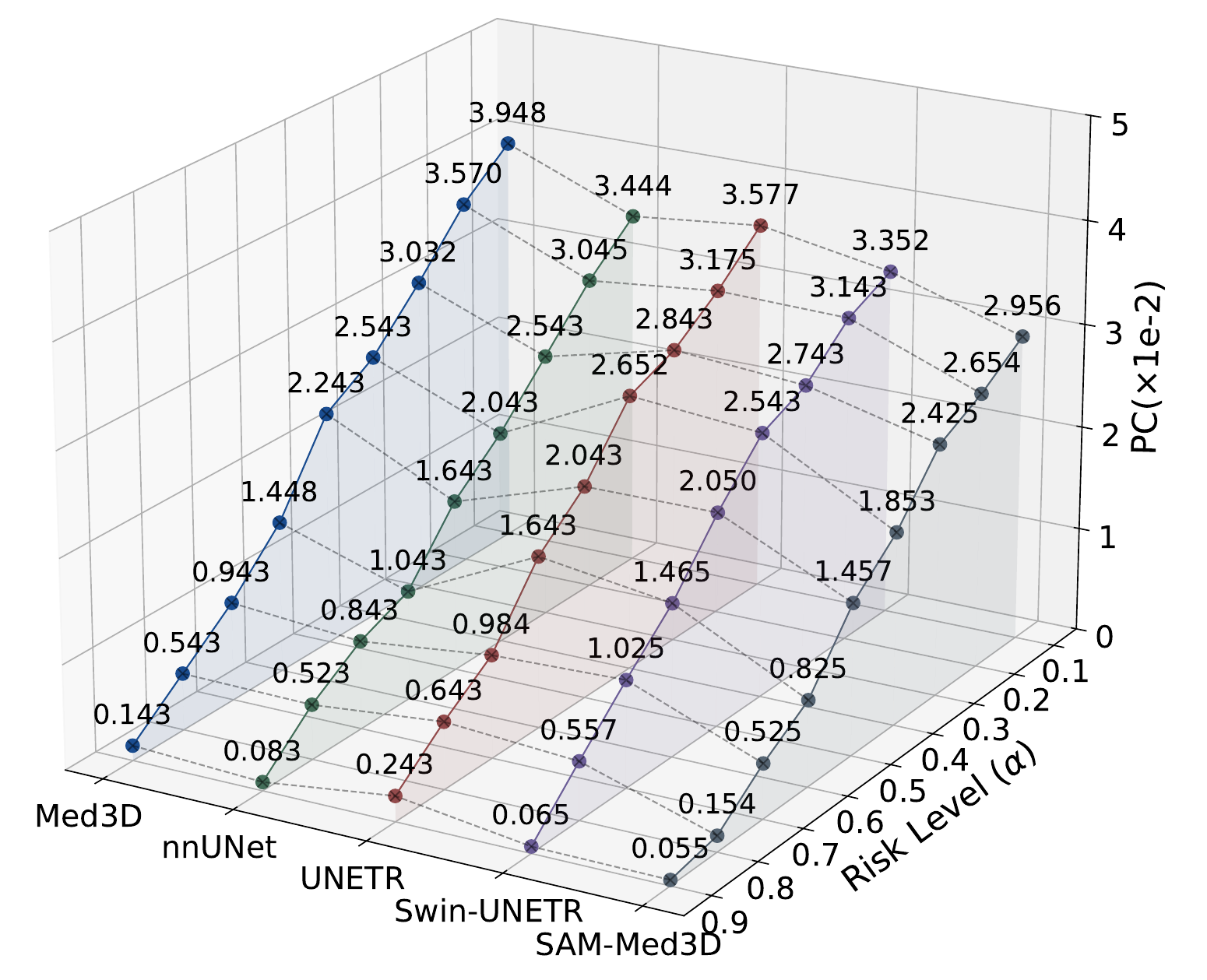}
    \caption{LIDC-IDRI.}
    \label{fig:APLVR results of LIDC-IDRI}
  \end{subfigure}
  \hfill
  \begin{subfigure}[b]{0.32\textwidth}
    \includegraphics[width=\textwidth]{img/cropped_APLVR_KITS21.pdf}
    \caption{MDSC-Colon.}
    \label{fig:APLVR results of MDSC-Colon}
  \end{subfigure}
  \hfill
  \begin{subfigure}[b]{0.32\textwidth}
    \includegraphics[width=\textwidth]{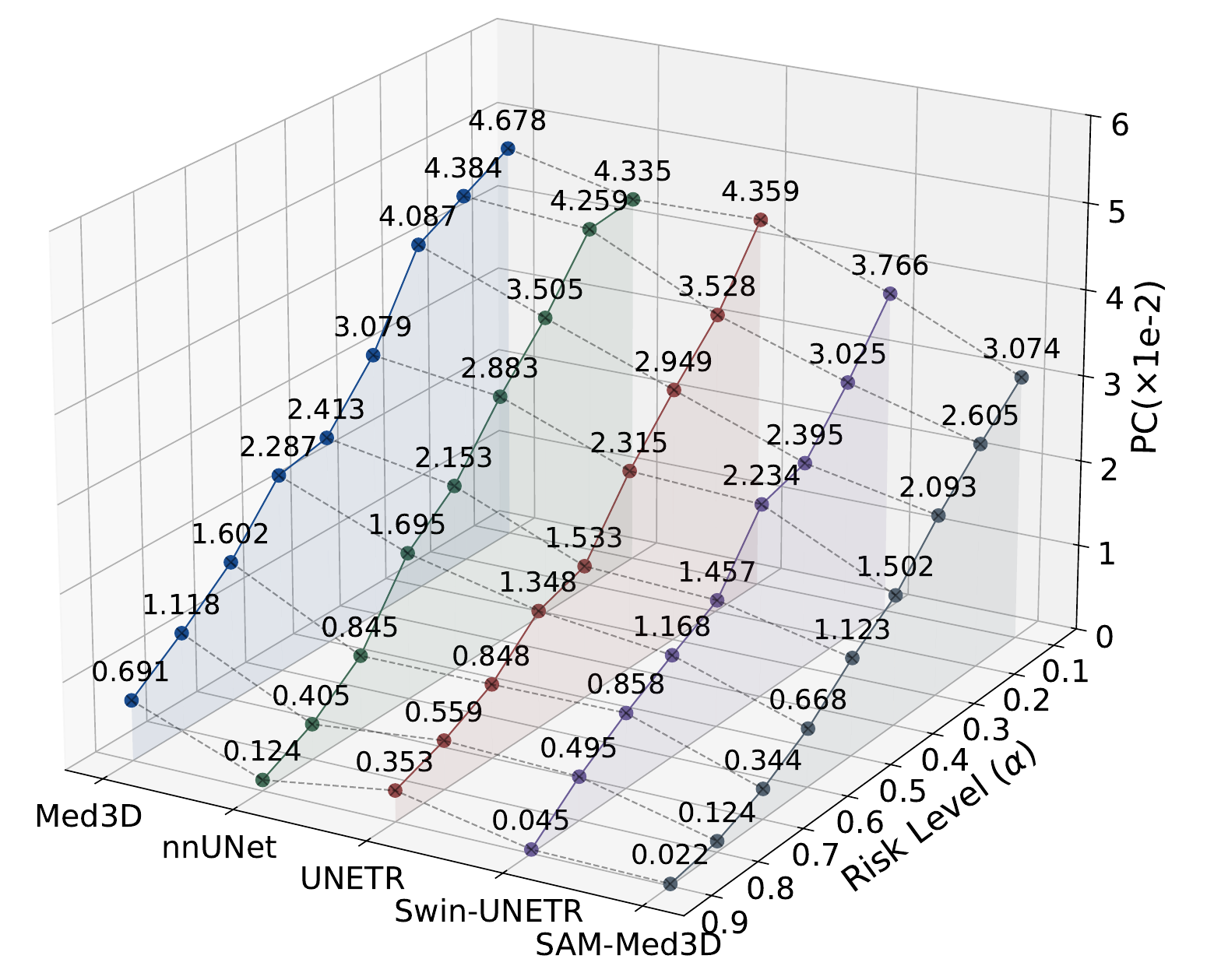}
    \caption{MDSC-Pancreas.}
    \label{fig:APLVR results of MDSC-Pancreas}
  \end{subfigure}

  \caption{PC results on six 3D-LS datasets.}
  \label{fig:APLVR}
\end{figure*}

\begin{figure}[t]
\centering
\includegraphics[width=\linewidth]{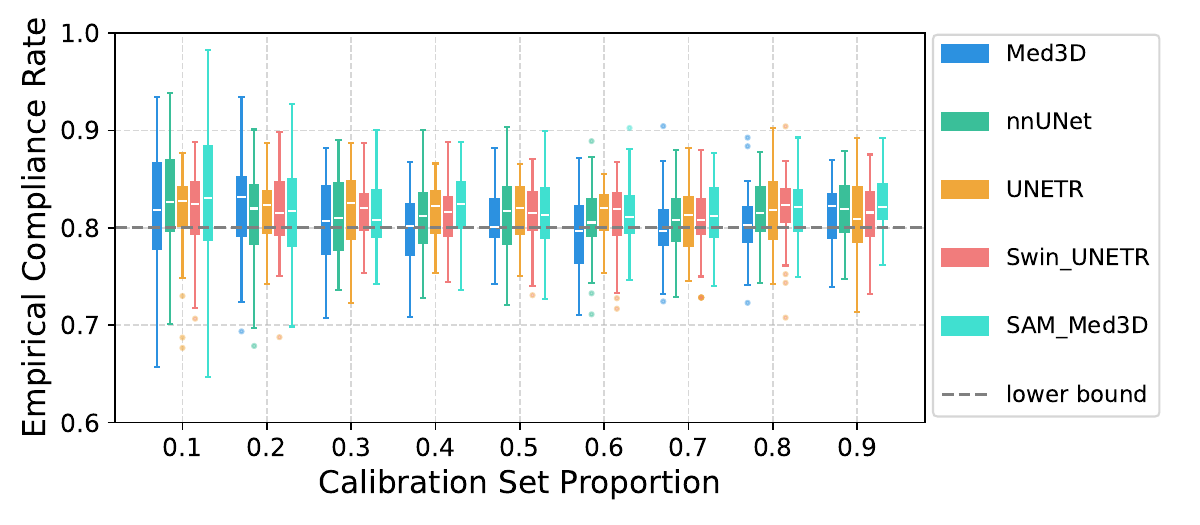}
\caption{ECR results using Med3D on KiTS21 under various calibration-to-test splits, with $\varepsilon = 0.4$ and $\alpha=0.2$.}
\label{fig:proportion}
\end{figure}

\noindent \textbf{Prediction Compactness as a Risk-aware Benchmark.} 
To further characterize model performance beyond FNR control, we examine PC across six datasets, comparing five segmentation models under varying risk levels. 
As illustrated in Figure~\ref{fig:APLVR}, the variation of PC with respect to the risk level $\alpha$ reveals how each model handles uncertainty. For most models, PC decreases monotonically with increasing $\alpha$, consistent with the intuition that lower risk levels allow for more aggressive (i.e., spatially expansive) lesion delineation.
This compactness improvement is not achieved at the expense of coverage: by construction, all results satisfy the predefined FNR tolerance $\varepsilon$, confirming that the reduction in PC reflects improved spatial precision under reliable conditions. In clinical segmentation tasks, such compact predictions are crucial for reducing false alarms and annotation overhead. 
Importantly, PC curves provide a model-agnostic benchmark for assessing the uncertainty structure and calibration efficiency. Models exhibiting a flatter PC curve (i.e., less sensitive to increases in $\alpha$) tend to allocate predictions more conservatively, indicating higher confidence concentration. 
By contrast, models with steep PC growth may reflect less calibrated uncertainty handling. 
These findings collectively demonstrate that CLS enables not only statistically guaranteed FNR control, but also introduces a robust, interpretable spatial metric—prediction compactness—under risk-aware evaluation. This highlights the utility of CLS in benchmarking and improving model uncertainty behavior in medical segmentation.

\noindent \textbf{Robustness to Calibration-Test Split Ratios.} 
To assess the robustness of CLS under varying calibration data availability, we evaluate its performance using different calibration-test split ratios. 
As shown in Figure~\ref{fig:proportion}, CLS maintains strict control of the test-time ECR across all configurations, even when only $10\%$ of the data is reserved for calibration. 
These results highlight a critical advantage of our method: statistical guarantees remain valid even in highly imbalanced calibration settings—a desirable trait for scalable and trustworthy deployment of risk-aware medical AI systems.  

\section{Conclusion}
In this paper, we present CLS, a risk-controlled lesion segmentation framework, which constructs a novel nonconformity score based on a tailored FNR-specific loss and establishes statistically rigorous decision thresholds via conformal calibration. 
Unlike heuristic thresholding, CLS consistently enforces test-time FNR control across diverse 3D-LS benchmarks, thereby substantially reducing the under-segmentation risk, which is an essential requirement for practical, safety-critical clinical applications. 
Beyond reliability, we introduce the \emph{prediction compactness} metric, serving as a novel, interpretable benchmark to quantify spatial precision and model uncertainty under formal risk constraints. 
We further demonstrate that CLS remains robust and effective even with limited calibration samples, supporting its applicability in resource-constrained settings. 
Overall, CLS offers a principled, flexible, and practical foundation for deploying uncertainty-aware segmentation models with statistical guarantees—bridging the gap between theoretical soundness and real-world clinical reliability.


\bibliography{aaai2026}

\newpage
\clearpage

\section{Appendix A: Additional Related Work}
\textbf{\textit{Background of Split Conformal Prediction.}}
We provide a detailed introduction to the standard split conformal prediction (SCP) procedure in classification tasks~\cite{angelopoulos2021gentle,wang2025sconu}. 
SCP provides a principled framework for transforming any heuristic or model-dependent notion of uncertainty into a statistically rigorous one. 
Given a held-out calibration set of size $N$, we compute the nonconformity score (NS) for each data point as one minus the softmax probability assigned to its ground-truth class. 
These scores are then sorted in ascending order, and we select the $\lceil (N + 1)(1 - \alpha) \rceil / N$ quantile as the threshold.

For a new test instance, we evaluate the softmax outputs across all classes. Any class whose softmax probability exceeds the derived threshold is included in the prediction set. Under the assumption of exchangeability, this construction guarantees that the true label will be included in the prediction set with approximate probability $1 - \alpha$. This procedure yields valid marginal coverage on finite-sample test data, offering statistically meaningful uncertainty estimates. The complete framework is presented as follows:

\begin{enumerate}
    \item 
    
Given the calibration dataset $\{(X_i, Y_i^*)\}_{i=1}^n$ (i.i.d.) and a pretrained model $\hat{f}(\cdot)$ that produces probabilistic predictions ($\hat{f}(X_i) \in [0,1]^K$, representing a probability distribution over $K$ classes for input $X_i$). The predicted probability assigned to the ground-truth class $Y_i^*$ is denoted by $\hat{f}(X_i)_{Y_i^*}$.

    \item 
    
Define the nonconformity score for each calibration sample as a measure of uncertainty associated with its true class: $s_i = s(X_i, Y_i^*) = 1 - \hat{f}(X_i)_{Y_i^*}$, where $\hat{f}(X_i)_{Y_i^*}$ denotes the predicted probability for the ground-truth label ($\{s_{1} \leq s_{2} \leq \cdots \leq s_{n}\}$).

    \item 

Compute the $\frac{\left \lceil (n+1)(1-\alpha ) \right \rceil }{n} $ quantile of $\{s_{i} \}_{i=1}^{n}$: $\hat{q}=\text{inf}\left \{ q:\frac{\left | \{i:s_i\le q \}\right | }{n} \ge \frac{\lceil (n+1)(1-\alpha )  \rceil}{ n}  \right \}=s_{\left \lceil (n+1)(1-\alpha ) \right \rceil }.$

    \item 

Construct the prediction set for $X_{{test}}$: $\mathcal{C}(X_{test}) = \left\{ y \in [K] : s(X_{\mathrm{test}}, y) \leq \hat{q} \right\}.$

    \item 

The event $Y^*_{\mathrm{test}} \in \mathcal{C}(X_{\mathrm{test}})$ is equivalent to the condition $s(X_{\mathrm{test}}, Y^*_{\mathrm{test}}) \leq \hat{q}$. As long as this inequality holds, the true label is guaranteed to be included in the prediction set $\mathcal{C}(X_{\mathrm{test}})$. Consequently, it can obtain a prediction set that successfully captures the ground-truth class.

    \item 

Owing to the exchangeability of the $n + 1$ data points (the $n$ calibration samples and one test instance), it has: $\mathbb{P}(s_{\mathrm{test}} \leq s_{(i)}) = \frac{i}{n + 1}.$

    \item 
Based on the exchangeability assumption, it can obtain that the probability of the prediction set covering the true label satisfies: $\mathbb{P}\left(Y^*_{\mathrm{test}} \in \mathcal{C}(X_{\mathrm{test}})\right) = \mathbb{P}\left(s_{\mathrm{test}} \leq \hat{q} \right) = \frac{\lceil (n + 1)(1 - \alpha) \rceil}{n + 1} \geq 1 - \alpha.$ This guarantees a marginal coverage level of at least $1 - \alpha$ on the test distribution.

\end{enumerate}

\section{Appendix B: Assumptions}
\textbf{\textit{Exchangeable data distribution.}}
We assume that the inputs to the segmentation model are independently and identically drawn from a fixed underlying distribution. This assumption is reasonable for many standard scenarios, such as the 3D medical imaging tasks for lesion segmentation explored in our experiments. However, it is important to note that this assumption does not hold in the presence of distribution shifts between the calibration and testing phases.

Throughout the paper, we utilize the following formal definition of exchangeable data distribution~\cite{farinhas2023non,wang2025sconu}, which is a weaker assumption than independent and identically distributed (i.i.d.) data. Let $\mathcal{X}$ and $\mathcal{Y}$ denote the input and output spaces, respectively. A data distribution in $\mathcal{X} \times \mathcal{Y}$ is said to be exchangeable if and only if $P\left((X_1, Y_1), \ldots, (X_n, Y_n)\right) = P\left((X_{\pi(1)}, Y_{\pi(1)}), \ldots, (X_{\pi(n)}, Y_{\pi(n)})\right)$ holds for any finite collection ${(X_i, Y_i)}_{i=1}^n \subseteq \mathcal{X} \times \mathcal{Y}$ and for any permutation $\pi$ of ${1, \ldots, n}$. It is important to note that every i.i.d. (independent and identically distributed) sequence is also exchangeable, since $P\left((X_1, Y_1), \ldots, (X_n, Y_n)\right) = \prod_{i=1}^n P(X_i, Y_i)$.

The proposed method exhibits potential for extension to settings involving certain types of distribution shift, which are common in real-world applications due to variations in data acquisition protocols, patient populations, or imaging modalities. One possible direction involves adapting the conformal calibration procedure by incorporating instance-wise or domain-adaptive weighting schemes when computing the nonconformity scores. These weighting schemes can be carefully designed to ensure that the resulting t-values retain the super-uniformity property relative to the target distribution, thereby preserving the validity guarantees under distributional changes and enhancing the robustness of the method in practical deployment.

\section{Appendix C: CLS Algorithm Description}
The algorithm begins with a calibration dataset $\mathcal{D}_{cal} = {(x_i, y_i^*)}_{i=1}^{n}$, where each $x_i$ denotes an input image and $y_i^*$ is the corresponding ground-truth segmentation mask. This calibration set is obtained as one realization from 100 random data splits to enhance the robustness and reliability of the evaluation. For each calibration sample, we employ a pretrained segmentation model $f(\cdot)$ to compute the corresponding confidence map $\hat{y}_i = f(x_i) \in [0, 1]^{D \times H \times W}$.

For each calibration sample, we perform a binary search to identify the smallest threshold $t_i$ such that the corresponding FNR-specific loss satisfies $\mathrm{L}_i^{\text{FNR}}(t_i) \le \varepsilon$, within a specified numerical tolerance $\delta$ (e.g., $10^{-4}$). This process yields a set of thresholds $\{t_i\}_{i=1}^{n}$, where each $t_i$ is individually calibrated to ensure that the user-defined false negative tolerance $\varepsilon$ is satisfied for its respective sample.

Subsequently, we compute the $\frac{\left \lceil (1-\alpha )(1+n) \right \rceil }{n}$ quantile of the sorted threshold set as the global decision threshold $\hat{t}$, which is then applied to segment unseen test instances. This guarantees, under the conformal prediction framework, that the resulting segmentation satisfies the false negative risk constraint $\Pr(\mathrm{L}^{\text{FNR}}_{test}(\hat{t}) \le \varepsilon) \ge 1 - \alpha$.

\begin{algorithm}[t]
\caption{\textsc{Conformal Lesion Segmentation}}
\label{alg:cls}
\begin{algorithmic}[1]
\REQUIRE \parbox[t]{0.95\linewidth}{
Calibration set $\mathcal{D}_{cal} = \{(x_i, y_i^*)\}_{i=1}^{n}$\\
Pretrained segmentation model $f(\cdot)$\\
FNR-specific loss $\mathrm{L}_i^{\text{FNR}}(t) = 1- \frac{| \mathrm{Mask}_i(t) \cap y_i^* |}{| y_i^* |}$\\
User-specified FNR tolerance $\varepsilon \in (0, 1)$ \\
Risk level $\alpha \in (0,1)$ \\
Fresh test instances $x_{test}$
}

\ENSURE 
\parbox[t]{0.95\linewidth}{
Segmentation mask $\mathrm{Mask}_{test}(\hat{t})$ \\
such that $\Pr\left( \mathrm{L}^{\text{FNR}}_{test}(\hat{t}) \leq \varepsilon \right) \geq 1 - \alpha$
}
\vspace{0.3em}
\FOR{$i = 1$ to $n$}
    \STATE $\hat{y}_i \gets f(x_i)$ 
    \STATE Initialize $t_{\min} \gets 0$, $t_{\max} \gets 1$
    \STATE Set numerical tolerance $\delta$ (e.g., $1e\!-\!4$)
    
\WHILE{$|\mathrm{L}_i^{\text{FNR}}(t) - \varepsilon| > \delta$}
    \STATE $t \gets \frac{t_{\min} + t_{\max}}{2}$
    \STATE $\mathrm{Mask}_i(t) \gets \left\{ (d,h,w) \in \Omega_{x_i} : \hat{y}_{i(d,h,w)} \ge 1 - t \right\}$
    \STATE $\mathrm{L}_i^{\text{FNR}}(t) \gets 1 - \frac{|\mathrm{Mask}_i(t) \cap y_i^*|}{|y_i^*|}$
    \IF{$\mathrm{L}_i^{\text{FNR}}(t) > \varepsilon$}
        \STATE $t_{\min} \gets t$
    \ELSE
        \STATE $t_{\max} \gets t$
    \ENDIF
    \STATE $t_i \gets t$
\ENDWHILE

\ENDFOR
\vspace{0.3em}
\STATE Sort thresholds $\{t_i\}_{i=1}^{n}$ in ascending order
\STATE $\hat{t} \gets \inf\left \{ t:\frac{\left |\{ i:t_i\le t \}\right | }{n} \ge \frac{\lceil (1-\alpha )(1+n)  \rceil}{n}  \right \}$
\STATE $\begin{aligned}
\mathrm{Mask}_{test}(\hat{t}) \gets \left\{ \substack{(d,h,w) \in \Omega_{x_{test}} : \\ f(x_{test})_{(d,h,w)} \ge 1 - \hat{t}} \right\}  
\end{aligned}$

\RETURN $\mathrm{Mask}_{test}(\hat{t})$
\end{algorithmic}
\end{algorithm}

\section{Appendix D: Details of Experimental Settings}
\subsection{D.1 Details of Datasets}
\input{tables/datasets}
As shown in the Table \ref{tab:datasets}, we employ six publicly available 3D medical image segmentation datasets, each targeting a distinct organ type. The column \#Series reports the number of annotated 3D imaging volumes (i.e., patient-level scans), \#Lesions indicates the total number of labeled lesion instances, and \#Train denotes the number of training samples utilized in our experimental setup.

\noindent\textbf{KiTS21} \cite{heller2023kits21} is a benchmark dataset released as part of a public challenge designed to advance automatic segmentation of kidneys and renal tumors from clinical abdominal CT scans. The dataset comprises multi-institutional CT volumes, each annotated independently by three experts, and includes a held-out test set from an external center to rigorously evaluate model generalizability. The dataset offers high-quality manual annotations for both kidney parenchyma and tumors, enabling robust and standardized performance comparisons across methods. Due to the substantial variability in tumor size, shape, and anatomical location, the KiTS21 dataset poses a challenging segmentation task and is widely utilized as a benchmark for evaluating the generalization capability of models in jointly segmenting organs and associated lesions.

\noindent\textbf{LiTS} \cite{bilic2023liver} is a widely adopted benchmark dataset for liver and tumor segmentation and detection, comprising 201 contrast-enhanced abdominal CT volumes collected from multiple clinical institutions. Each volume is annotated with pixel-wise labels for the liver and intrahepatic tumors, including both primary and secondary lesions. The dataset poses significant challenges due to the low contrast, ambiguous boundaries, and diverse morphological characteristics of the tumors. Despite these complexities, LiTS remains a standard benchmark in the field and is particularly valuable for assessing model performance on small lesion segmentation and hepatic pathology analysis.

\noindent\textbf{NIH-LN ABD} \cite{roth2014new} is a benchmark dataset developed by the National Institutes of Health (NIH) for evaluating lymph node detection and multi-organ segmentation algorithms. It consists of abdominal CT scans with pixel-level annotations for multiple organs, including the liver, spleen, kidneys, pancreas, and abdominal lymph nodes. The dataset is particularly challenging due to the low contrast, variable size of lymph nodes, and the anatomical complexity of abdominal structures. NIH-LN ABD is widely used in research involving lymph node detection, classification, and multi-organ segmentation in clinically complex contexts.

\noindent\textbf{LIDC-IDRI} \cite{armato2011lung} is a widely used public dataset for the development and evaluation of computer-aided detection and diagnosis systems for pulmonary nodules. It consists of 1018 low-dose thoracic CT scans, each annotated by four experienced radiologists using a two-phase review process (initially blinded, then unblinded). The dataset includes a total of 7371 marked lesions, with 2669 nodules annotated with detailed contours and subjective characteristics. Following prior studies \cite{de2025uls23}, we selected a subset of 2236 nodules for our analysis, based on the availability of complete annotations and clinical relevance. Rich metadata such as nodule size, location, and boundary definition makes LIDC-IDRI a valuable resource for research on lung nodule segmentation, particularly in the context of inter-observer variability and uncertainty-aware modeling.

\noindent\textbf{MDSC-Colon} \cite{antonelli2022medical} is a subtask of the Medical Segmentation Decathlon (MSD), specifically curated to evaluate the effectiveness of segmentation algorithms in colorectal tumor analysis. This dataset consists of contrast-enhanced abdominal CT scans collected from colon cancer patients across multiple clinical institutions, accompanied by high-quality, expert-annotated tumor delineations. The segmentation task is notably challenging due to the substantial heterogeneity in tumor morphology, including irregular shapes, asymmetry, and the presence of diffuse or poorly defined boundaries. MDSC-Colon serves as a rigorous benchmark for assessing a model’s ability to handle anatomically complex and variable lesion presentations.

\noindent\textbf{MDSC-Pancreas} \cite{antonelli2022medical} represents another task within the Medical Segmentation Decathlon (MSD), aimed at evaluating segmentation performance on the pancreas and its associated pathological structures, including tumors and cysts. The dataset comprises contrast-enhanced abdominal CT scans acquired from multiple clinical institutions, each annotated with high-precision labels for both the pancreas and relevant lesions. Pancreatic segmentation poses considerable challenges due to the organ’s small size, highly variable shape, and low contrast relative to surrounding anatomical structures.

\subsection{D.2 Details of Models and Fine-tuning Strategies}
\input{tables/models}
We selected five 3D medical image segmentation models, each characterized by a distinct architectural design, and applied customized fine-tuning strategies to align their parameter counts for a fair and consistent comparison. Table~\ref{tab:models} provides a summary of the selected models and their respective parameter counts following the fine-tuning strategies.

\noindent\textbf{Med3D} \cite{chen2019med3d} is a convolutional neural network (CNN) framework based on an encoder-decoder architecture, specifically adapted for 3D medical image segmentation. Its encoder is derived from the ResNet family and modified to accommodate volumetric data. To ensure a fair comparison with transformer-based models, Med3D increases the number of channels and convolutional layers at each stage, thereby expanding the model’s capacity while maintaining its fully convolutional design. These enhancements enable Med3D to serve as a strong CNN baseline for evaluating downstream 3D segmentation performance.

\noindent\textbf{nnU-Net} \cite{isensee2021nnu} is a self-configuring, CNN-based segmentation framework built upon the standard U-Net encoder-decoder architecture. Designed to adapt to the characteristics of each target dataset, nnU-Net automatically determines optimal preprocessing steps, network configurations, and training protocols. In this work, we follow prior practice by increasing the number of filters and deepening the convolutional blocks at each resolution level, thereby matching the model’s parameter count with transformer-based counterparts. These adaptations ensure a fair comparison in downstream segmentation performance.

\noindent\textbf{UNETR} \cite{hatamizadeh2022unetr} integrates Vision Transformers (ViTs) into the classical U-shaped architecture for 3D medical image segmentation, leveraging transformers’ ability to model long-range dependencies while preserving the spatial precision of convolutional decoders. Although convolutional layers excel at capturing local features, they often struggle with global semantic understanding \cite{xiao2023transformers}. To enable a fair comparison with CNN-based models, UNETR is modified by reducing the hidden size, MLP dimensionality, and number of attention heads in the transformer encoder. These modifications substantially decrease the parameter count while maintaining the core advantages of transformer-based representation learning.

\noindent\textbf{Swin-UNETR} \cite{hatamizadeh2021swin} integrates a Swin Transformer-based encoder with a fully convolutional decoder for 3D medical image segmentation. It employs a hierarchical architecture with shifted window self-attention, enabling efficient modeling of both local and global features while significantly reducing computational overhead. To maintain a comparable parameter scale with other models, the network’s depth and feature dimensions are deliberately reduced. These modifications retain the core advantages of window-based attention, while enhancing scalability and reducing the overall computational burden.

\noindent\textbf{SAM-Med3D} \cite{wang2023sam} extends the Segment Anything Model (SAM) framework to 3D medical image segmentation by adapting its core components, including the image encoder, prompt encoder, and mask decoder, to volumetric data. The model employs 3D convolutions, learnable 3D absolute positional embeddings, and 3D attention mechanisms to support spatial representation in three dimensions. To reduce model complexity, SAM-Med3D decreases the embedding dimensions and the number of attention heads in the transformer layers. This lightweight design maintains the original ViT-based architecture while improving efficiency and performance in 3D medical segmentation tasks.

\end{document}

%% file: tables/average-FNR-specific-loss.tex
\begin{table*}[!t]
\caption{Test-time FNR (mean ± std) on the KITS21 dataset under fixed versus CLS-calibrated decision thresholds at varying FNR tolerance levels ($\varepsilon$).}
\centering
\begin{tabular}{cccccc}
\toprule
\textbf{$\varepsilon$ / Models}                 
                         & \textbf{Med3D} & \textbf{nnUNet} & \textbf{UNETR} & \textbf{Swin-UNETR} & \textbf{SAM-Med3D} \\ 
\midrule        
\multicolumn{6}{c}{$t=0.5$ (Fixed by default)}                                                  \\ 
\midrule
{}            &    0.5338$\pm$0.2018   &    0.5122$\pm$0.1971    &      0.5804$\pm$0.2110     &    0.5642$\pm$0.1571   &         0.4935$\pm$0.1478   \\ \midrule
\multicolumn{6}{c}{$t=\hat{t}$ calibrated at the risk level of 0.2}                                             \\ 
\midrule
0.1                   &   0.0533$\pm$0.0169    &   0.0649$\pm$0.0201    &    0.0576$\pm$0.0196     &   0.0466$\pm$0.0136    &     0.0469$\pm$0.0165       \\
0.2                     &   0.1032$\pm$0.0212    &    0.0933$\pm$0.0238    &      0.1145$\pm$0.0263     &   0.1070$\pm$0.0290    &        0.0956$\pm$0.0286    \\
0.3               &   0.1654$\pm$0.0413    &    0.1764$\pm$0.0346    &      0.1693$\pm$0.0267     &   0.1501$\pm$0.0377    &     0.1779$\pm$0.0328       \\
0.4                &    0.2253$\pm$0.0836   &   0.2378$\pm$0.0753     &     0.2201$\pm$0.0523      &    0.2540$\pm$0.0688   &       0.2473$\pm$0.0623     \\
0.5               &     0.4211$\pm$0.0501  &    0.4345$\pm$0.0679    &     0.3928$\pm$0.0540      &    0.4022$\pm$0.0617   &         0.3895$\pm$0.0669   \\
\bottomrule
\end{tabular}
\label{tab:ablation thresholding mean}
\end{table*}

%% file: tables/average-ECR.tex
\begin{table*}[ht]
\caption{Comparison of test-time ECR (mean ± std) on six 3D-LS benchmarks using heuristic (fixed $t=0.5$) and CLS-calibrated thresholds under a risk level of $\alpha = 0.2$.}
\centering
\begin{tabular}{cccccc}
\toprule
\textbf{Datasets / Models}                 
                         & \textbf{Med3D} & \textbf{nnUNet} & \textbf{UNETR} & \textbf{Swin-UNETR} & \textbf{SAM-Med3D} \\ 
\midrule        
\multicolumn{6}{c}{$t=0.5$ (Fixed by default)}                                                  \\
\midrule
KITS21                   &    0.4117$\pm$0.0454  &    0.4072$\pm$0.0368    &     0.4549$\pm$0.0378      &   0.4081$\pm$0.0351    &      0.5623$\pm$0.0432      \\
LITS                     &   0.3635$\pm$0.0206    &   0.3384$\pm$0.0283     &      0.3877$\pm$0.0187     &    0.3134$\pm$0.0289    &        0.6002$\pm$0.0210    \\
NIH-LN ABD               &   0.4275$\pm$0.0322    &    0.4156$\pm$0.0294    &      0.4311$\pm$0.0243     &   0.4078$\pm$0.0305    &        0.5975$\pm$0.0372    \\
LIDC-IDRI                &    0.4783$\pm$0.0466   &     0.4802$\pm$0.0458   &      0.4512$\pm$0.03964     &    0.4233$\pm$0.0478   &       0.5788$\pm$0.0401     \\
MDSC-Colon               &   0.3029$\pm$0.0261    &   0.3489$\pm$0.0297     &    0.3677$\pm$0.0313       &    0.3167$\pm$0.0212   &         0.5499$\pm$0.0397   \\
MDSC-Pancreas            &    0.4588$\pm$0.0453   &    0.4725$\pm$0.0376    &      0.3935$\pm$0.0428     &    0.4360$\pm$0.0451   &         0.6277$\pm$0.0380   \\ \midrule
\multicolumn{6}{c}{$t=\hat{t}$ calibrated at the risk level of 0.2}\\                                             \midrule
KITS21                   &   0.8080$\pm$0.0513    &   0.8027$\pm$0.0454    &    0.8041$\pm$0.0506     &   0.8289$\pm$0.0429    &     0.8031$\pm$0.0523       \\
LITS                     &   0.7988$\pm$0.0387    &    0.8103$\pm$0.0393    &      0.8086$\pm$0.0278     &   0.8147$\pm$0.0322    &        0.8104$\pm$0.0404    \\
NIH-LN ABD               &   0.8156$\pm$0.0227    &    0.8198$\pm$0.0275    &      0.8024$\pm$0.0301     &   0.8078$\pm$0.0337    &     0.8254$\pm$0.0375       \\
LIDC-IDRI                &    0.8137$\pm$0.0523   &   0.8006$\pm$0.0502     &     0.7969$\pm$0.0496      &    0.8125$\pm$0.0433   &       0.8094$\pm$0.0511     \\
MDSC-Colon               &     0.8034$\pm$0.0378  &    0.8023$\pm$0.0343    &     0.8165$\pm$0.0376      &    0.8166$\pm$0.0424   &         0.8154$\pm$0.0344   \\
MDSC-Pancreas            &    0.8012$\pm$0.0243   &    0.7943$\pm$0.0298    &     0.8034$\pm$0.0220      &     0.8104$\pm$0.0322  &         0.8060$\pm$0.0348   \\ 
\bottomrule
\end{tabular}
\label{tab:ablation thresholding p}
\end{table*}

%% file: tables/datasets.tex
\begin{table}[!t]
\caption{Summary statistics of 3D-LS datasets.}
\footnotesize 
\setlength{\tabcolsep}{4pt} 
\begin{tabular}{c c c c c} 
\toprule
\textbf{Datasets} & \textbf{Data Type} & \textbf{\#Series} & \textbf{\#Lesions} & \textbf{\#Train} \\ \midrule 
KiTS21            & Kidney             & 300               & 332                & 249              \\
LiTS              & Liver              & 113               & 832                & 624              \\
NIH-LN ABD        & Lymph Nodes        & 85                & 557                & 417              \\
LIDC-IDRI         & Lung               & 750               & 2236               & 1677             \\
MDSC-Colon        & Colon              & 126               & 131                & 98               \\
MDSC-Pancreas     & Pancreas           & 281               & 283                & 212              \\ 
\bottomrule
\end{tabular}
\label{tab:datasets}
\end{table}

%% file: tables/models.tex
\begin{table}[!t]
\caption{Models and parameter sizes.}
\centering
\setlength{\tabcolsep}{4pt} 
\begin{tabular}{c c c c c} 
\toprule
\textbf{Models} & \textbf{Parameters (M)}  \\ 
\midrule
Med3D            & 43.8365                   \\
nnUNet              & 44.2345                    \\
UNETR        & 44.6862              \\
Swin-UNETR         & 45.0454                     \\
SAM-Med3D        & 45.5447                    \\
\bottomrule
\end{tabular}
\label{tab:models}
\end{table}